\documentclass[prb,twocolumn,english,showpacs]{revtex4}
\usepackage{graphicx}
\usepackage{amssymb}

\begin{document}

\title{A trapped single ion inside a Bose-Einstein condensate}

\author{Christoph Zipkes, Stefan Palzer, Carlo Sias, and Michael K{\"o}hl}

\affiliation{Cavendish Laboratory, University of Cambridge, JJ Thomson Avenue, Cambridge CB3 0HE, United Kingdom}

\maketitle

\textbf{Improved control of the motional and internal quantum states of ultracold neutral atoms and ions has opened intriguing possibilities for quantum simulation and quantum computation. Many-body effects have been explored with hundreds of thousands of quantum-degenerate neutral atoms\cite{Bloch2008} and coherent light-matter interfaces have been built\cite{Brennecke2007,Colombe2007}. Systems of single or a few trapped ions have been used to demonstrate universal quantum computing algorithms\cite{Blatt2008} and to detect variations of fundamental constants in precision atomic clocks\cite{Rosenband2008}.Until now, atomic quantum gases and single trapped ions have been treated separately in experiments. Here we investigate whether they can be advantageously combined into one hybrid system, by exploring the immersion of a single trapped ion into a Bose-Einstein condensate of neutral atoms. We demonstrate independent control over the two components within the hybrid system, study the fundamental interaction processes and observe sympathetic cooling of the single ion by the condensate. Our experiment calls for further research into the possibility of using this technique for the continuous cooling of quantum computers\cite{Daley2004}. We also anticipate that it will lead to explorations of entanglement in hybrid quantum systems and to fundamental studies of the decoherence of a single, locally controlled impurity particle coupled to a quantum environment\cite{Leggett1987,Recati2005}.}

Hybrid systems allow the exploration of physics much more advanced than that which can be studied using the component systems alone. In particular, the immersion of distinguishable particles into a quantum liquid has contributed significantly to our understanding of many-body systems. In liquid helium, for example, vortex lattices have been observed using charged particles as markers for the vortex lines\cite{Yarmchuk1979}. Moreover, in conventional\cite{Abrikosov1961,Yazdani1997} and high-Tc\cite{Pan2000} superconductors single impurity atoms lead to quasiparticle excitations which profoundly affect the superconducting properties. In future investigations of distinct single particles in combination with quantum matter, it will be important to have a high degree of control over these particles. In this regard, the excellent control possible over the external and internal degrees of freedom of a single ion trapped in a Paul trap is highly promising. Immersed into a quantum gas, a single trapped ion not only acts as a probe but could also be used for local manipulation. Numerous applications of this hybrid system have been foreseen, including sympathetic cooling\cite{Makarov2003}, the nucleation of localized density fluctuations in a Bose gas\cite{Cote2002,Massignan2005,Goold2009}, scanning probe microscopy with previously unattainable spatial resolution\cite{Kollath2007,Sherkunov2009}, and hybrid atom-ion quantum processors\cite{Idziaszek2007}. The majority of these proposals are based on there being a strong collisional interaction between the ion and the neutral atoms.

The interaction between a single ion and a neutral atom is dominated by the polarization potential. Asymptotically this behaves as\cite{Cote2000} $V(r)=-\frac{1}{(4\pi\epsilon_0)^2 }\frac{\alpha q^2}{2 r^4}$ and therefore decays more slowly than the van der Waals interaction between two neutral atoms. Here $r$ is the internuclear separation, $q$ denotes the charge of the ion, $\alpha$ is the dc polarizability of the neutral atom, and $\epsilon_0$ is the vacuum permittivity. The characteristic radius of the interaction potential is $r^*=\sqrt{\frac{\mu \alpha q^2}{(4\pi\hbar\epsilon_0)^2}}$ in which $\mu$ denotes the reduced mass of the two collision partners. For example, for neutral $^{87}$Rb atoms and $^{174}$Yb$^+$ ions $r^*=271$\,nm. In the s-wave scattering regime we expect the elastic scattering cross section to be on the order of $\sigma_0\approx 4\pi r^{*2}$. Due to the slow decay of the polarization potential, pure s-wave scattering can be reached only at collision energies in the neV energy range \cite{Cote2000,Idziaszek2009}. At higher collision energies $E \gg \frac{\hbar^2}{\mu r^{*2}}$, when several partial waves contribute to scattering, the cross section is still determined by $r^*$ and behaves\cite{Cote2000} $\sigma(E)=\pi(1+\pi^2/16) r^{*2} \left(\frac{\hbar^2}{\mu r^{*2} E}\right)^{1/3}$.

One particularly valuable application of the strong atom-ion interaction is sympathetic cooling of the ion using ultracold atoms. The possibility of cooling objects by immersing them in an ultracold atomic gas has been discussed theoretically for ions\cite{Makarov2003}, polar molecules\cite{Soldan2004}, and nanomechanical resonators\cite{Treutlein2007}. The atomic gas provides a reservoir which is much colder than the motional energy scales of the trapped ion and, if elastic collisions dominate, sympathetic cooling\cite{Larson1986} can be achieved.

\begin{figure}
 \includegraphics[width=\columnwidth,clip=true]{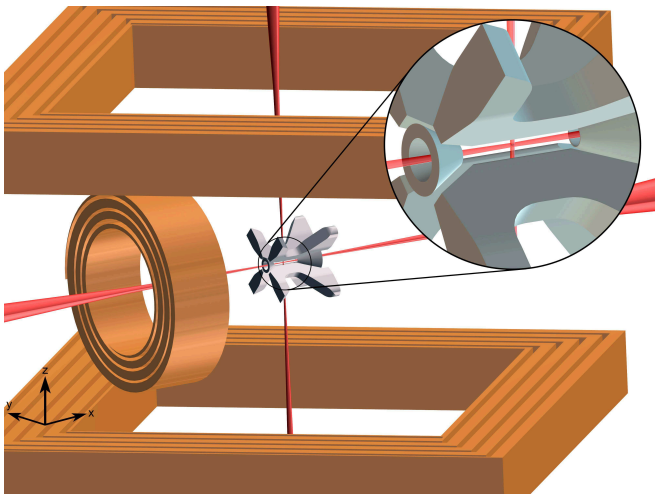}
 \caption{Experimental apparatus. The linear Paul trap (grey) is sandwiched in between the solenoids (copper) used for magnetically trapping the neutral atoms. Initially the neutral atoms are prepared outside of the ion trap, and subsequently moved into the ion trap through a bore in the endcap electrode. In the final position, the atoms are loaded into a crossed-beam optical dipole trap (red) in which the final evaporative cooling to Bose-Einstein condensation is performed (see inset). The drawing is to scale and the distance between the square magnetic field coils is 36\,mm.}
 \label{fig1}
\end{figure}

We prepare a single trapped ion inside a Bose-Einstein condensate by combining an ion trap and magnetic and optical traps for the neutral atoms into one setup (see Figure 1 and Methods). We employ a linear Paul trap to trap a single $^{174}$Yb$^+$ ion. The neutral atoms are prepared\cite{Palzer2009}, using standard techniques, as an ultracold but initially non-degenerate gas of neutral $^{87}$Rb atoms in the $|F=2,m_F=2\rangle$ hyperfine ground state in a magnetic trap, located 8\,mm away from the center of the ion trap. Subsequently, we magnetically transport the neutral atoms into the ion trapping region and prepare a Bose-Einstein condensate of approximately $3\times10^4$ atoms in an optical dipole trap several tens of $\mu$m away from the trapped ion.

During the preparation cycle of the neutral atoms (magneto-optical trapping, magnetic transport and evaporation in the magnetic trap), which lasts for $~60 s$, and before interaction with the ultracold neutral cloud, the ion is not laser-cooled. Owing to resonant heating from the radio-frequency radiation used in evaporative cooling of the neutral atoms, the ion reaches a temperature of $27.5 \pm 0.6$\,K, as measured by laser fluorescence (Methods). During preparation of the Bose-Einstein condensate in the optical dipole trap, the ion initially partly overlaps with the large thermal cloud of neutral atoms at 200\,nK, for 2\,s. After this interaction, the temperature of the ion is measured by laser fluorescence to be $T < 0.5$\,K, a reading whose precision is limited only by the resolution of the fluorescence measurement technique. This signals sympathetic cooling of the ion by the neutral atom cloud.

\begin{figure}
\includegraphics[width=\columnwidth]{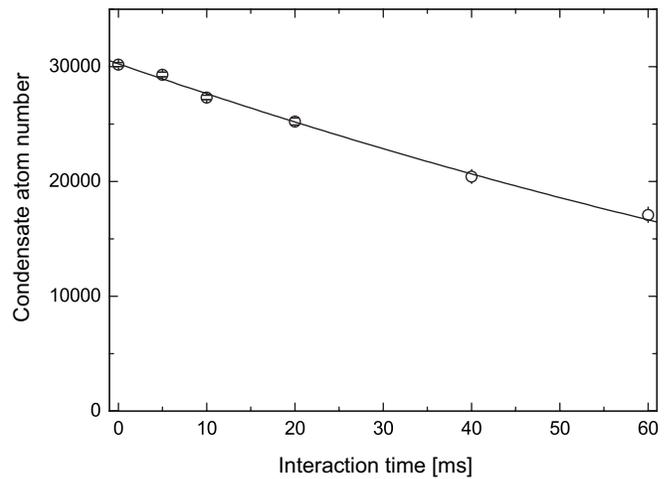}
 \caption{Atom number loss of a Bose-Einstein condensate due to collisions with a single ion. The solid line is a fit using a theoretical model (see Methods) and is used to determine the cross section $\sigma_{al}$ for neutral atom loss. Each data point is averaged over approximately 40 repetitions of the experiment and the standard error is given. The bare atom loss rate without the ion present has been subtracted.}
 \label{fig2}
\end{figure}

In order to observe interactions between an ion and a Bose-Einstein condensate we quickly move the pre-cooled ion over a distance of $\Delta x=140\,\mu$m into the center of the Bose-Einstein condensate. The potential depth of the optical dipole trap is  1\,$\mu$K which is less than the vibrational level spacing of the ion trap potential. Therefore, every collision will result in a neutral atom loss. We monitor this loss as a function of the interaction time. The data are displayed in Figure 2. From the decay of the atom number we estimate the cross section for atom loss to be $\sigma_{al}=(2.2\pm0.2)\times10^{-13}$\,m$^{2}$ (see Methods). The measured atom loss cross section allows us to estimate the collision energy between the ion and the neutral atoms and thus the ion temperature more precisely. An ion trapped in a harmonic oscillator of frequency $\omega$ at a temperature $T$ has an rms spread in position of $\sqrt{k_B T/(m_{ion} \omega_{ion}^2)}$. Convolving the position spread with the theoretical expression for the elastic scattering cross section $\sigma(E=k_BT)$ and equating to the measured result of $\sigma_{al}$ suggests $T=(2.1\pm 0.5)$\,mK. This temperature is comparable to the Doppler temperature limit of laser cooling of $T_D=0.5$\,mK. Buffer gas cooling in the simplified case of a classical gas of hard-spheres has been predicted\cite{Major1968} to reach temperatures of $k_BT \approx m_{ion} \Omega^2 d^2/2$ in which $d$ is the amplitude of the micromotion in the Paul trap. Comparing the theoretical estimate with our data suggests a micromotion amplitude of approximately 2\,nm which is compatible with our trap parameters. Extrapolating to the limit in which $d\rightarrow 0$ leads to $k_BT \approx \hbar \omega$, which could enable ground-state cooling.

We have time-resolved the dynamics of the immersion cooling process by immersing a hot ion directly into the Bose-Einstein condensate. For this measurement we partially suppress precooling of the ion in the neutral thermal atom cloud by displacing the ion along a diagonal direction in the x-y-plane. As a result, the initial temperature of the ion is approximately 4\,K. Then we quickly move the ion into the Bose-Einstein condensate, wait for a variable interaction time, release the neutral atoms, and measure the ion's temperature using the fluorescence method (see Figure 3a). The fluorescence method, although limited in its temperature resolution, is a reliable and independent method in this temperature range. We observe sympathetic cooling of the ion in the condensate on a timescale of a few tens of milliseconds. After 60 ms, the ion has reached a temperature of T$ = 0.6 \pm 0.7$\,K, corresponding to a temperature as low as our resolution limit of this measurement. We simultaneously monitor the atom loss rate (Fig. 3b). While the ion is hot, atom losses are very small and we estimate a cooling efficiency of approximately 1,000 vibrational quanta per collision. This is very efficient in comparison with laser-cooling, in which typically only a few quanta are removed per collision. Furthermore, while the ion is hot the radius of its trajectory is comparable to the size of the condensate. As the ion cools, it becomes more strongly localized at the centre of the condensate and the loss rate of atoms increases, possibly as a result also of the increase in the collisional cross-section, $\sigma(E)$

\begin{figure}
\includegraphics[width=\columnwidth]{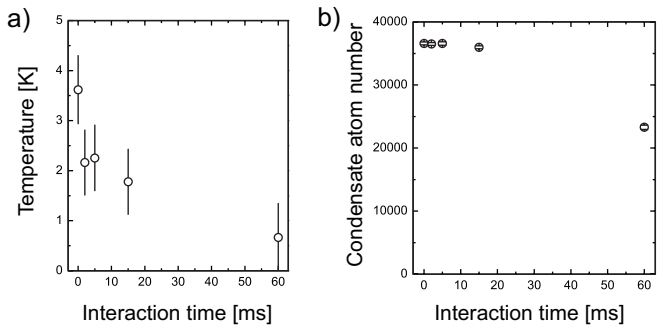}
 \caption{Time-resolved sympathetic cooling of a single ion. a) Temperature decrease as a function of the interaction time. The temperature has been measured by the laser fluorescence method and the standard deviation of the maximum likelihood estimate is shown. b) Atom loss during sympathetic cooling. While the ion is hot atom losses are small and the cooling rate approached 1000 vibrational quanta per collision. When the ion cools it localizes inside the region of higher density and the atom loss rate increases. The standard errors are shown. Each data point is averaged over approximately 200 repetitions of the experiment.}
 \label{fig4}
\end{figure}

Apart from elastic scattering, there are other possible interaction processes between an ion tightly confined in a Paul trap and an ultracold neutral atom. Firstly, atoms and ions can undergo charge exchange collisions\cite{Cote2000,Makarov2003,Idziaszek2009,Grier2009} of the type Yb$^+$+Rb$\rightarrow$Yb+Rb$^+$ or by creating a charged molecule Yb$^+$+Rb$\rightarrow$YbRb$^+$. This process is signaled by a loss of the Yb$^+$ ion from the ion trap. The choice of our elements aims at suppressing charge exchange by making the process off-resonant. Secondly, there are inelastic collisions due to micromotion of the ion. Collisions in the presence of this driven oscillation can result in an energy transfer from the driving electric field to the colliding partners.

A small charge exchange collision rate is indicated by a low loss probability of the ion in collisions with the neutral atoms. The ion loss probability as a function of the interaction time is shown in Figure 4a, using the same experimental sequence as for Figure 2. We observe a linear increase of the ion loss probability with the interaction time and find that charge exchange is suppressed with respect to elastic collisions by a factor of $6\times 10^{-6}$. From this we estimate a loss rate constant of  $K_{ce}=(1.4\pm0.5)\times 10^{-20}$m$^3$/s, five orders of magnitude smaller than in the homonuclear case\cite{Grier2009} owing to the large energy gap of 2 eV between the collision channels. The bare ion loss probability in the measurement interval is $10^{-4}$ which we derive from our ion trap lifetime of 8\,minutes. The results clearly signal the suppression of charge exchange processes for our choice of elements.

The effect of micromotion on ultracold atom-ion collisions has not been deeply explored so far. For the classical collision regime, theoretical considerations of micromotion-induced collisions exist\cite{Major1968,Devoe2009} and its influence on charge exchange processes has been seen\cite{Grier2009}. We investigate micromotion effects experimentally by applying a dc electric field $E_{offset}$ to the ion. The electric field slightly displaces the ion from the geometric center of the trap and introduces an excess micromotion of amplitude $d\approx 2.3$\,nm$\times E_{offset}$/(V/m). An ultracold atomic cloud is then overlapped with the ion and we study atom trap loss. This set of measurements has been performed with a thermal atom cloud of approximately $2\times 10^5$ atoms at 200\,nK trapped in a magnetic trap with frequencies $\omega_{\perp,mag}=2\pi\times 28$\,Hz and $\omega_{x,mag}=2\pi\times 8$\,Hz. We center the position of the magnetic trap onto the position of the displaced ion by applying homogeneous magnetic offset fields for each value of $E_{offset}$. For the actual measurement we quickly move the ion to the center of the atomic cloud, hold it there for 8 seconds, and measure the neutral atom loss rate from the cloud. Figure 3b shows our data for different values of the offset electric field. We find a clear signature of the effects of inelastic collisions due to micromotion leading to larger atom loss rates for increased micromotion with $|E_{offset}|>5$\,V/m. We attribute the shape of the curve to the energy dependence of atom-ion scattering rate which scales $\propto |E_{offset}|^{1/3}$.

\begin{figure}
\includegraphics[width=\columnwidth]{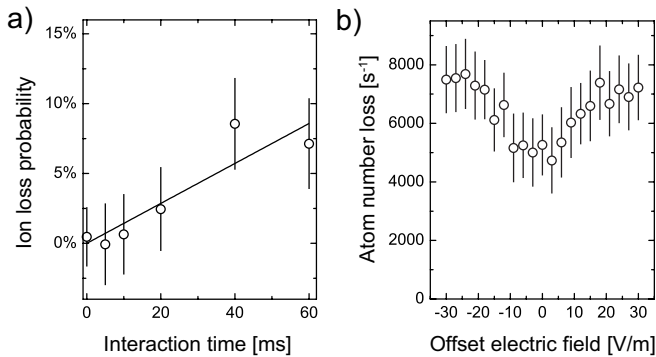}
 \caption{Inelastic atom-ion collisions. {\bf a)} Loss probability for a single ion inside a Bose-Einstein condensate. The time-dependent part of the loss is attributed to interaction processes between atoms and ions. We determine an upper bound to the charge exchange collision rate from a linear fit to the data. Every data point is averaged over approximately 170 repetitions of the experiment. {\bf b)} Micromotion induced atom losses for a single ion in a magnetically trapped thermal cloud. The data are averaged over 40 repetitions of the experiment for each data point and for two different orientations of the offset electric field. The standard error is given.}
 \label{fig3}
\end{figure}

Our results highlight the possibilities for sympathetic cooling of a single ion using an ultracold atomic bath to temperatures that are relevant for quantum information processing. Cooling qubits by immersion into a superfluid has been predicted to maintain their quantum coherence and entanglement6. Therefore, atomic Bose-Einstein condensates may serve as ideal refrigerators for ion-trap quantum computers. The successful immersion of a single trapped ion in a Bose-Einstein condensate makes several new experiments possible. The high degree of spatial control of a single ion facilitates the manipulation and study of quantum-degenerate samples with nanometre precision. Resolving single sites in a three-dimensional optical lattice is within reach\cite{Kollath2007,Sherkunov2009} and even the creation of entanglement between distant lattice sites seems feasible. The atom-ion interaction is predicted to be tunable by means of atom-ion Feshbach resonances\cite{Idziaszek2009}. This will allow fundamental studies of systems comprising a single impurity in a bath, for example the spin-boson model \cite{Leggett1987} or the bosonic version of Anderson's orthogonality catastrophe\cite{Sun2004}.

We are grateful to N. Cooper, C. Kollath, D. Lucas, D. Moehring, E. Peik, and C. Wunderlich for discussions. We acknowledge support from EPSRC, ERC (Grant number 240335), and the Herchel Smith Fund (C.S.). Correspondence and requests for materials should be addressed to M.K.~(email: mk540@cam.ac.uk).

All authors contributed to the design, data acquisition, and interpretation of the presented work. C.Z. and S.P.
 contributed equally to the construction of the apparatus and to the acquisition of the data.

\section*{Methods summary}
A single $^{174}$Yb$^+$ ion is stored in a radio-frequency Paul trap which creates a three-dimensional harmonic oscillator potential of characteristic frequencies $\omega_{\perp,ion}= 2\pi\times 2\cdot 10^5$\,Hz and $\omega_{x,ion}= 2\pi\times 5\cdot 10^4$\,Hz. The ion is cooled and probed by near-resonant laser light. The laser light is detuned by $\Delta=-\Gamma/2$ from the $S_{1/2} \rightarrow P_{1/2}$ transition at a wavelength of 370\,nm. $\Gamma=2 \pi \times 20$\,MHz is the linewidth of the atomic transition. The intensity of the light scattered from the ion is sensitive to the ion's velocity which is used to determine the temperature. An ultracold cloud of neutral $^{87}$Rb atoms in the $|F=2,m_F=2\rangle$ hyperfine ground state is moved into the ion trapping volume using a time-varying inhomogeneous magnetic field. There, Bose-Einstein condensates of up to $3\times 10^4$ atoms are prepared by transferring the neutral atoms into a far-detuned optical dipole trap formed by two crossed laser beams at 935\,nm wavelength and performing forced evaporation by continuously lowering the trap depth. The final trap frequencies of the optical trap are $\omega_\textrm{x,opt}=2\pi\times 51$\,Hz, $\omega_\textrm{y,opt}=2\pi\times 144$\,Hz, $\omega_\textrm{z,opt}=2\pi\times 135$\,Hz.

\section*{Methods}

\subsection{Ion trapping}
The ion is confined in the ponderomotive potential (pseudo-potential) of a linear Paul trap which is created from a very rapidly oscillating electric quadrupole field. The trap operates at an RF frequency of $\Omega=2\pi\times 42.7$\,MHz, well above the typical frequencies used in evaporative cooling of the neutral atoms. The distance between the ion and the RF electrodes is $R=0.5$\,mm and the applied voltage of $V_{RF}\approx 250$\,V gives rise to a radial confinement of $\omega_{\perp,ion}= 2\pi\times 2\cdot 10^5$\,Hz. The axial confinement is $\omega_{x,ion}= 2\pi\times 5\cdot 10^4$\,Hz. A single $^{174}$Yb$^+$ ion is loaded into the ion trap by isotope-selective two-photon ionization\cite{Balzer2007} from a pulsed neutral Ytterbium atomic beam. Subsequently the ion is laser cooled on the $S_{1/2}\rightarrow\, P_{1/2}$ transition at 370 nm. Repumping light is provided at 935 nm to clear population out of the metastable $D_{3/2}$ state. We continually cancel residual dc electric fields, which would result in excess micromotion of the ion, to better than 0.5\,V/m using a photon-correlation technique\cite{Berkeland1998}. Using the same technique we establish an upper bound of 7\,nm of uncompensated micromotion of the ion along the symmetry axis of the linear trap.

\subsection{Preparation of Bose-Einstein condensates}
We transport the cold, but non-degenerate, neutral atomic cloud from its initial position into the ion trap by displacing the potential minimum of the magnetic Ioffe trap using suitably timed changes of the electrical currents in the solenoids. The neutral atoms enter the ion trap through a 700\,$\mu$m wide bore in the end cap electrode, moving along the symmetry axis of the linear ion trap. Shortly before the neutral atoms arrive at their final position we displace the ion by 140\,$\mu$m from its ideal position. We then load the neutral atoms into an optical dipole trap formed by two crossed laser beams operating at 935\,nm wavelength centered onto the ideal position of the ion. Each of them is focused to a waist of approximately 70\,$\mu$m and powered by up to a few hundred milliwatts. In the dipole trap we perform evaporative cooling by lowering the laser intensity to reach Bose-Einstein condensates of $3\times 10^4$ atoms. The trap frequencies of the optical trap are $\omega_\textrm{x,opt}=2\pi\times 51$\,Hz, $\omega_\textrm{y,opt}=2\pi\times 144$\,Hz, $\omega_\textrm{z,opt}=2\pi\times 135$\,Hz. We apply a homogeneous bias magnetic field of less than 1\,G to prevent depolarization of the neutral atomic sample. The cross-talk between the trapping potentials for atoms and ions is measurable but very small. The electric quadrupole field of the ion trap provides a potential for the neutral atoms via the dc Stark effect. This gives rise to a weak anti-trapping harmonic potential of approximately 10 Hz frequency which does not significantly interfere with atom trapping. In the time-of-flight images of the neutral atoms we observe this anti-trapping potential as an additional acceleration during expansion of the cloud. The effect of the optical trapping potential onto the trapped ion is considerably weaker and can be neglected.

\subsection{Temperature measurements of the ion}
We measure the temperature of the ion using a laser fluorescence technique\cite{Epstein2007,Wesenberg2007}. When illuminated with a near-resonant laser beam a cold ion displays a high level of fluorescence whereas a hot ion will exhibit smaller fluorescence due to the large Doppler shift. During the detection the temperature of the ion will change due to Doppler laser cooling and the fluorescence will increase to its saturation value over a time span of several hundred microseconds. We detect the time-resolved fluorescence of the ion on a photomultiplier tube using a binning time of 25\,$\mu$s. The frequency of the detection laser is red-detuned by $\Delta=-\Gamma/2$ from the $S_{1/2}\rightarrow P_{1/2}$ transition of $^{174}$Yb$^+$ ($\Gamma=2\pi\times 20$\,MHz denotes the linewidth of the atomic transition). We analyze the temporal increase of the fluorescence by numerically calculating the solution of the time-dependent optical Bloch equations and performing a thermal ensemble average\cite{Wesenberg2007}. The resulting temperature is deduced by a $\chi^2$-fit to the time-resolved data. For an ion laser cooled to the Doppler temperature of $T_D=0.5$\,mK this measurement results in a temperature of $T=0.2\pm0.3$\,K, indistinguishable from an ion sympathetically cooled in a thermal bath, for which the fluorescence method yields $T=0.3\pm0.5$\,K. The relatively coarse temperature resolution of this method of a few hundred mK is determined by photon shot noise and the number of averages.

\subsection{Determination of the scattering cross section.}
We model the Bose-Einstein condensate with a density distribution $n(\vec{r})$ in the Thomas-Fermi approximation and with the ion pinned at the centre of the Bose-Einstein condensate ($\vec{r} = 0$). The cross section $\sigma_{al}$ for atom loss has contributions from the scattering cross section as well as from the spatial extend of the ion. The cold ion locally depletes the condensate and the characteristic velocity at which the condensate density readjusts is velocity of sound $c=\sqrt{4\pi\hbar^2 a_0 n(0)/m_{Rb}^2}$. Here $a_0=5.45$\,nm is the s-wave scattering length of the neutral Rb atoms and $m_{Rb}$ their mass. We assume that in every collision an atom is lost from the trap since the vibrational level spacing of the ion trap is larger than the potential depth of the optical dipole trap. In this model the atom loss obeys the rate equation
\begin{equation}
\frac{dN}{dt}=-\frac{15^{3/5}\overline{\omega}^{9/5} m_{Rb}^{4/5}}{\sqrt{8}\pi a_0^{2/5}\hbar^{4/5}}\,\sigma_{al}\,N(t)^{3/5}
\label{eqn1}
\end{equation}
in which $\overline{\omega}=(\omega_{x,opt}\omega_{y,opt}\omega_{z,opt})^{1/3}$. We fit the solution of equation (\ref{eqn1}) to our data to extract the scattering cross section $\sigma_{al}$.

\end{document}